\documentclass[prl,aps,floatfix,showpacs,showkeys,amsmath,amssymb,nofootinbib,twocolumn,10pt]{revtex4-1}
\usepackage{amsthm}
\usepackage{amsfonts}
\usepackage{amstext}
\usepackage{graphicx}
\usepackage{color}
\usepackage{enumitem}
\usepackage[dvipsnames]{xcolor}
\usepackage{tkz-kiviat,numprint} 
\usetikzlibrary{arrows}
\usetikzlibrary{plotmarks}
\usepackage{dcolumn}
\usepackage{booktabs}

\usetikzlibrary{shapes}
\usepackage{xspace}
\usepackage{mje}

\definecolor{mygreen}{RGB}{13,140,53}
\definecolor{myorange}{RGB}{220,60,30}
\definecolor{vincecolor}{RGB}{102,0,51}
\definecolor{markcolor}{RGB}{0,0,153}

\newcommand{\QSERG}{Quantum Systems Engineering Research Group, Loughborough University, Leicestershire LE11 3TU, United Kingdom}
\begin{document}


\title{On Calculating the Dynamics of Very Large Quantum Systems}

\author{J.J. Bowen}
\affiliation{\QSERG}
\author{V.M. Dwyer}
\affiliation{\QSERG}
\author{I.W. Phillips}
\affiliation{\QSERG}
\author{M.J. Everitt}\email{m.j.everitt@physics.org}
\affiliation{\QSERG}
\date{\today}

\begin{abstract} 
Due to the exponential growth of the state space of coupled quantum systems it is not possible, in general, to numerically store the state of a very large number of quantum systems within a classical computer. 
We demonstrate a method for modelling the dynamical behaviour of measurable quantities for very large numbers of interacting quantum systems. 
Our approach makes use of a symbolic non-commutative algebra engine that we have recently developed in conjunction with the well-known Ehrenfest theorem. Here we show the possibility of determining the dynamics of experimentally observable quantities, without approximation, for very large numbers of interacting harmonic oscillators. 
Our analysis removes a large number of significant constraints present in previous analysis of this example system (such as having no entanglement in the initial state).  
This method will be of value in simulating the operation of large quantum machines, emergent behaviour in quantum systems, open quantum systems and quantum chemistry to name but a few.
\end{abstract}

\maketitle

Obtaining information for large numbers of interacting quantum systems is in general a hard problem. 
This is due to the fact that the Hilbert/state space describing the total system grows exponentially in size with the number of its constituents. 
There are many open questions that arise from this behaviour in areas such as open quantum systems, emergent phenomena such as quantum phase transitions and the quantum to classical crossover.  
Beyond basic science, the emergence of new quantum technologies demands computer aided tools to serve a similar purpose to those, for example, in the semiconductor industry
(see ~\cite{Gough2009,combes2016slh,PhysRevA.81.023804,5286277})  and may be expected to range from simple quantum circuit modellers (a quantum analogue of Spice) to more sophisticated quantum VLSI applications. 
The programming languages most often deployed within the physical sciences are not well equipped to deal with the non-commuting algebra of quantum systems that such engineering tools would need to accommodate. 
We note that, while some basic functionality has recently been introduced into Maple, it is not yet well suited to the solution of complex problems.
For this reason we are developing a non-commuting symbolic algebra package using a functional programming language (Haskell) that we have named QuantAl (for Quantum Algebra). 
In this work we report our first application of QuantAl in the study of the exact dynamics of a very large set of interacting quantum systems.
We will discuss, in detail, the form and operation of this package in a dedicated follow-up paper. Here we note that it is based on using an $N$-ary tree data structure to represent mathematical expressions, using recursion to traverse through the tree, applying pattern matching to perform operations, and respecting non-commutativity of operators within the expressions.  
In the current work we apply this technique to enable the modelling and simulation of large interacting quantum systems, observing that in some cases this can be done without approximation. 
In addition to better understanding the physics of many body quantum systems there are a number of technological applications which readily arise from this process including: design for test; verification, certification and validation; system design and parameter estimation. 

In studying large numbers of interacting quantum systems, in the absence of analytical solutions, some compromise is always necessary. 
Often, for example, models are produced that provide approximate information for only part of the system (the reduced density operator from a master equation). 
In this work we take a different, complementary, approach and seek limited information on the complete dynamics, through monitoring the expected values of observable quantities. This greatly simplifies the analysis as it then becomes unnecessary to represent the quantum state at all. 
One possible approach begins by considering one of the oldest results  of quantum theory - Ehrenfest's Theorem~\cite{Ehrenfest1927}, and
then uses to advantage symbolic computer algebra to automate the process of generating the large, but often tractable, number of the equations of motion needed to determine the dynamics of the expectation values. 

The expectation value, for any quantum mechanical operator \OpA, is related to its commutation relation with a system's Hamiltonian \OpH according to~\cite{Ehrenfest1927}:
\bel{eq:ET}
\frac{\ud}{\ud t}\EX{\OpA} 
=
{-\frac{\ui}{\hbar}}\EX{\COM{\OpA}{\OpH}} +
\EX{\frac{\partial \OpA}{\partial t}}
\ee
For one dimensional systems of position \Opq and momentum \Opp, whose Hamiltonian with potential energy $V(\Opq)$, is 
\be
\OpH = \frac{\Opp^2}{2m} + V(\Opq)
\ee
Ehrenfest's theorem, when applied to \Opq and \Opp in turn, yields the following well-known  result
\be
{\frac {\ud}{\ud t}}\EX{\Opq} =\frac{\EX{\Opp}}{m} 
\mathrm{\ and\ }
{\frac {\ud}{\ud t}}\EX{\Opp} =
-\EX{\frac {\partial V(\Opq)}{\partial \Opq}}.
\ee
These two coupled equations are of the same form as the classical equation of motion for position and momentum of the same system. We also observe the obvious fact that these two equations form a closed set of first order differential equations. In 1998 Nicholas Wheeler discussed at length, in an unpublished essay, some useful consequences of the Ehrenfest's theorem including more complicated one-dimensional systems and a useful discussion on the momental hierarchy supported by arbitrary observables~\cite{Wheeler98}. Within our method any truncation of the momental hierarchy may introduce an approximation. 
If the hierarchy is finite, the method is exact; and in many cases will not suffer exponential scaling of the Hilbert space of composite systems. If the moments do not converge the availability of this semi-analytical method makes other options practically available such as the following. If we note that, using the evolution operator,
$$
\ME{\psi(t)}{\OpA}{\psi(t)}
=
\ME{\psi(0)}{e^{\frac{\ui \OpH t}{ \hbar}}\OpA e^{-\frac{\ui \OpH t}{ \hbar}}}{\psi(0)}.
$$
together with the Baker-Campbell-Hausdorff formula expansion
it is simple to form a power series solution 
\ba
\EX{\OpA}(t)=& \EX{\OpA}_{t=0}+\frac{\ui t}{\hbar} \EX{\left[ \OpH ,\OpA\right]}_{t=0}+ \nonumber \\ &
\frac{1}{2!}\left(\frac{\ui t}{\hbar}\right)^2\EX{[\OpH,[\OpH,\OpA]]}_{t=0} +\ldots
\ea 
which is accurate if computed to sufficiently high orders. This alternative method will be explored in a later publication. In this work we will make use of Ehrenfest's theorem for a system for which there are no momental hierarchy issues and for which the method is very efficient - namely that of a set of coupled simple harmonic oscillators.

Before proceeding to this example we outline the process of applying Ehrenfest's theorem.
To determine the equation of motion for $\EX{\OpA}$ we are required to not just apply~\Eq{eq:ET} once, but recursively to all those expectation values that arise from $\EX{\COM{\OpA}{\OpH}}$ and $\EX{\partial \OpA / \partial t}$.
We must continue this process, stopping only when a closed set of equations is formed or some sufficient approximation is achieved (or a sufficient momental hierarchy is established). 
Once this is done we will have obtained a set of first order ordinary differential equations that can be solved to compute system dynamics using standard tools such as Matlab, and thus
require only the systems initial conditions.
The setting of these is trivial when the state is separable; where only small subsets of its constituents are entangled, or when analytical methods can be deployed.
For example, expectation values of maximumly entangled coherent states are easy to determine so long as the system's observables may be represented in terms of normally ordered creation and annihilation operators. 
Given that time-dependent Hamiltonians do not necessarily greatly complicate this method, it should also be possible to include more complex states as initial conditions when they are known to arise as the outcome of another dynamical process, such as through gate operations.


Our example system comprises a set of harmonic oscillators coupled, in the usual way, via the position degrees of freedom. The total Hamiltonian in this case is 
\be
\OpH=\sum_{i=1}^N   \hbar \omega_i \left(\Opn_i + \Half \right)+ \sum_{i<j} \gamma_{ij} \Opq_i\Opq_j
\ee
where $\omega_i$ is the frequency of the $i^\mathrm{th}$ oscillator, $\Opn_i$ is its number operator, $\Opq_i$ its position operator and  $\gamma_{ij}$ the coupling between oscillators. 
We note that adding to this Hamiltonian time-dependent drive terms such as $\Opq_i \sin \Omega_i t$ or time-dependent and/or non-linear couplings such as $\gamma_{ij}(t) \Opq_i^m\Opq_j^n$ (with $m,n <4$)
\footnote{Issues begin to arise when quartic and higher powers are introduced as the momental hierarchy is infinite and thus requires truncation.} 
would not significantly complicate the computational analysis; a powerful feature of this method.
 
Let us single out, in this first example, one of the oscillators, say $i=0$, and arrange the only non-zero couplings to be $\gamma_{0i}=\Gamma, i=1,2,\ldots$. 
Under these circumstances, the Hamiltonian $\OpH$ describes the evolution of one simple harmonic oscillator ($i=0$) open to an environment (bath) comprising an infinite set of Harmonic Oscillator modes of frequency $\omega_i, i=1, 2, \cdots$, and coupling constants $\gamma_i$. 
This is a well studied system corresponding to Quantum Brownian Motion~\cite{Fleming2011}.  
Due to the high symmetry, this system is one where our application of the QuantAl package is not strictly necessary as it is possible to calculate the form of the equations of the momental expansion by hand. Nevertheless, it provides an excellent test as (i) our results are  easily checked for correct output and (ii) this system is one which is very familiar to those working in open quantum systems. 
In this work we concentrate on demonstrating that this proposed alternative to modelling dynamics of large quantum systems is an good one, more complex systems will be examined in later works that focus on specific interesting physics.
The influence of the bath on the system is typically characterised by a spectral density $J(\omega)$ which is formed of the density of the bath states (modes) weighted by the strength of system-mode couplings. 
A common set of simplifications used to make progress include the Born-Markov and Ohmic Bath approximations, which generally rely on a weak coupling between system and bath, assume that the time scales describing the system dynamics are much slower than those of the bath, removing memory effects, and assume the spectral density takes on the very simple form $J(\omega)=2\gamma \omega$, cut off by an often arbitrary frequency $\Omega$. 

\begin{figure}[!t]
\begin{tikzpicture}
        \path (0,0) node(a) {\includegraphics[width=0.9\linewidth,trim={4.7cm 10cm 4.4cm 10cm},clip]{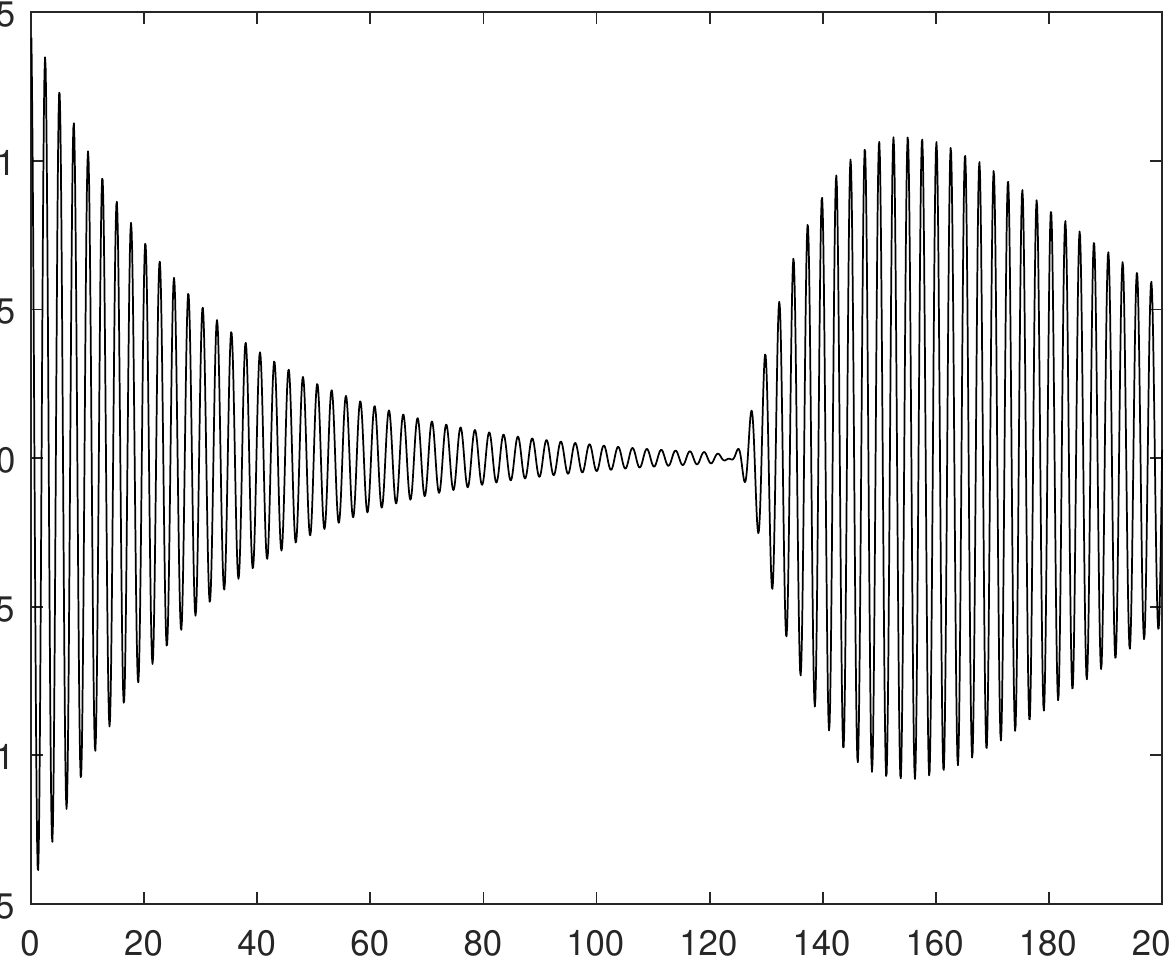}};
        \path (-0.3,-1.5) node(b) {\includegraphics[width=0.35\linewidth,trim={4.4cm 10cm 3.5cm 9.6cm},clip]{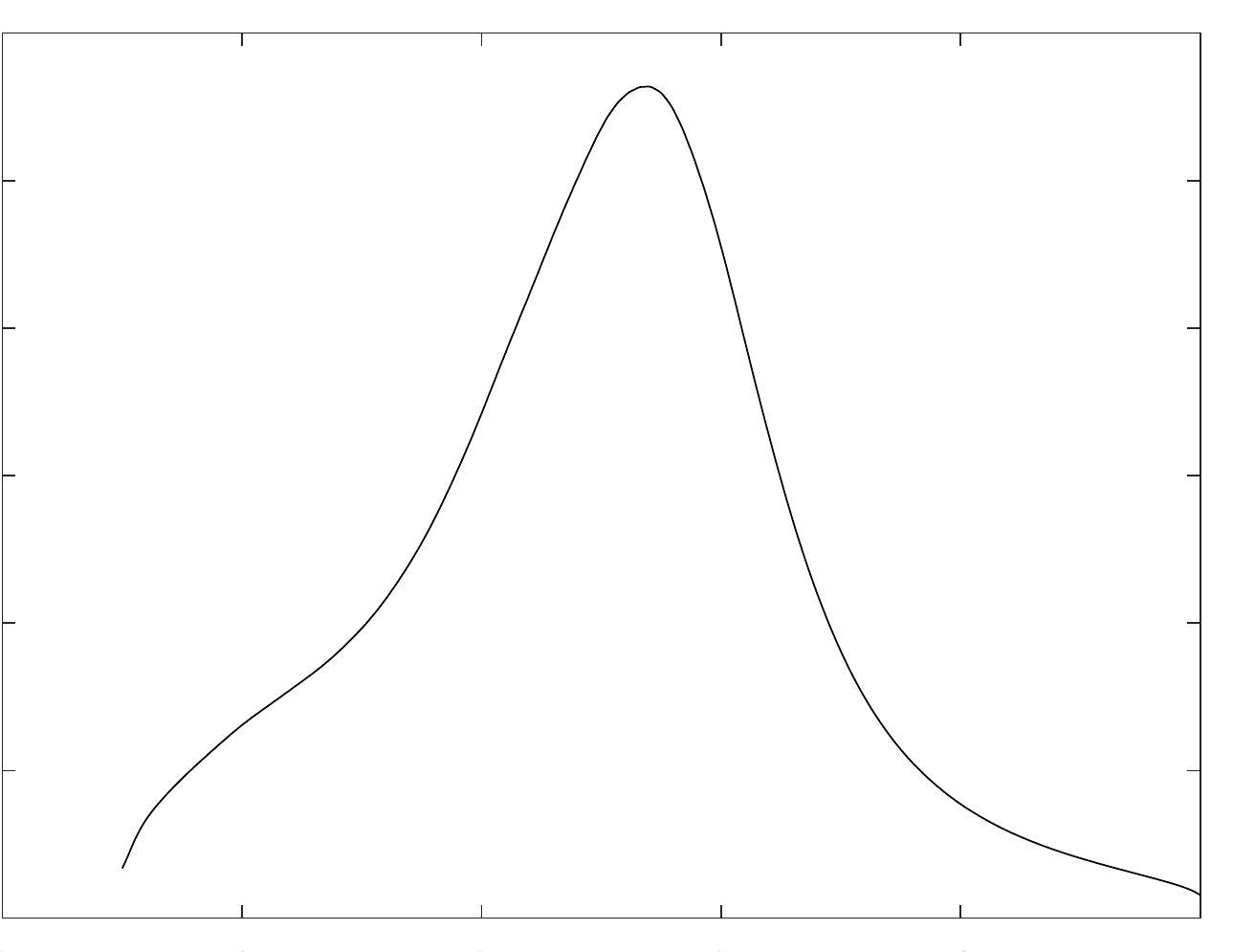}};
  \node[label=left:\rotatebox{90}{Dimensionless position \EX{\Opq}}] at (-4,0) {};
  \node[label=below:{Dimensionless time $\omega_0 t$}] at (0,-3.2) {};
  \node[label={\large \textsf{(a)}}] at (3.4,2.2) {};
\end{tikzpicture}    
\begin{tikzpicture}
        \path (0,0) node(a) {\includegraphics[width=0.9\linewidth,trim={4.7cm 10cm 4.4cm 10cm},clip]{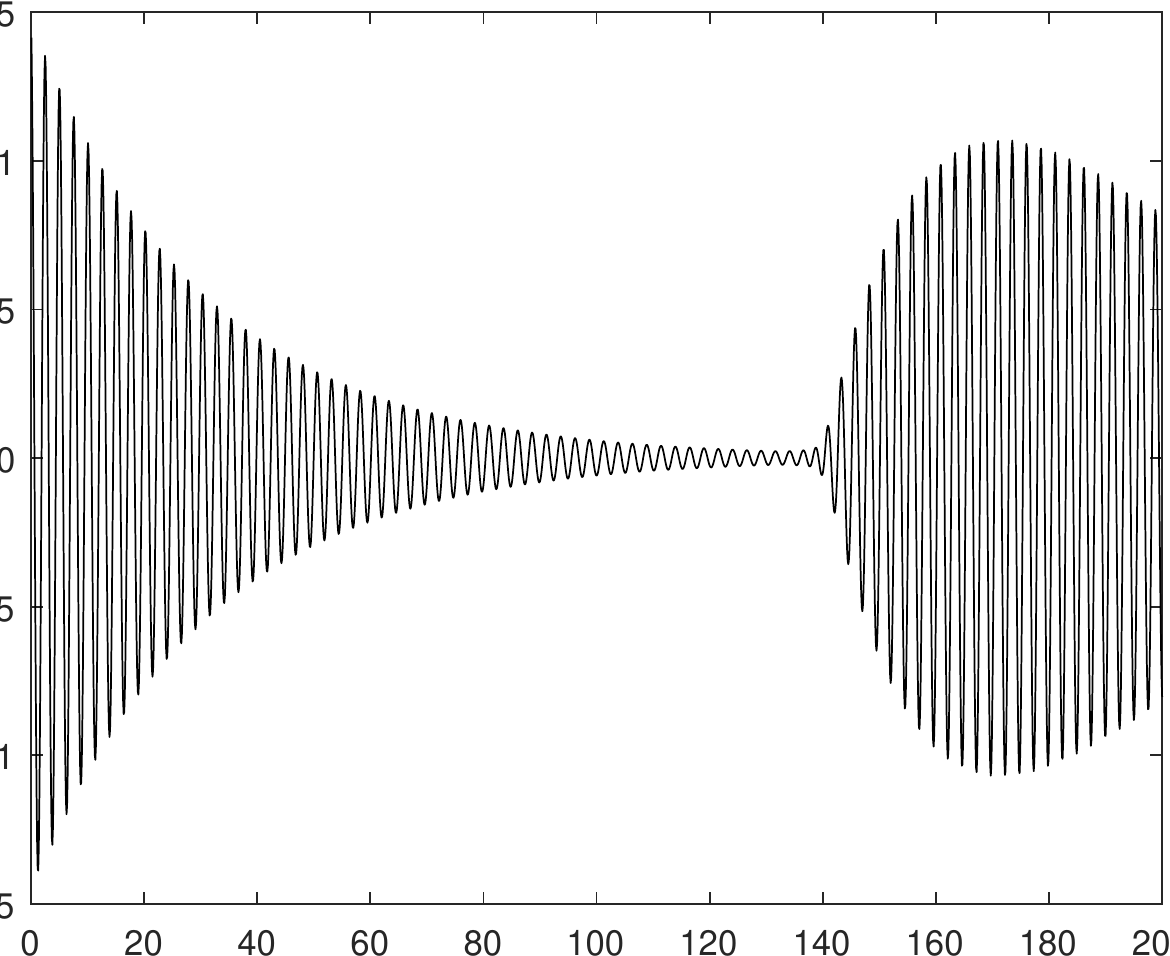}};
  \node[label=left:\rotatebox{90}{Dimensionless position \EX{\Opq}}] at (-4,0) {};
  \node[label=below:{Dimensionless time $\omega_0 t$}] at (0,-3.2) {};
  \node[label={\large \textsf{(b)}}] at (3.4,2.2) {};
  \draw[red, dashed, thick] (1,3) -- (1,-2.8);
\node[label=left:{\textsf{Revival time in (a)}}] at (1,-2.2) {};
\end{tikzpicture} 
    \caption{\label{figSeperable} Here we show the dynamics of the expectation value of position \EX{\Opq_0} for a simple harmonic oscillator staring in a coherent state \ket{\alpha=1} coupled to a bath of oscillators each staring in their own ground state \ket{\alpha=0}. The top figure (a) shows the oscillator's dynamics when the bath comprises 100 oscillators and the bottom plot (b) when it contains 200. In line with expectations, it is evident that a bath of 100 oscillators results in an earlier revival than for one comprising 200 oscillators. The inset in (a) shows the form of the spectral distribution of the bath that we have used to produce both plots.
    }
\end{figure}

There are a number of important cases in which such assumptions are unrealistic; including systems strongly coupled to their environment and importantly, for the development of quantum technologies, those systems which operate at low temperatures. Moreover experimentally determined spectral densities are seldom simple, their structure often being complicated by multiple peaks. 
Within this context, the present method allows for the study of non-Markovian evolution of an arbitrary system coupled to an environment of essentially arbitrary complexity. 
In~\Fig{figSeperable}(a) and (b) we compare the full quantum dynamics of the expectation value of position \EX{\Opq_0} for a bath comprising 100  and 200 oscillators respectively. 
Here~\Fig{figSeperable} shows ``damping'' and revival phenomena of the fully quantum bath comprising (a) 100 and (b) 200 oscillators. As expected, the revival time (associated with non-Markovian effects) occurs later in the case where the bath is larger. 

\begin{figure}[!t]
\begin{tikzpicture}
        \path (0,0) node(a) {\includegraphics[width=0.9\linewidth,trim={4.7cm 10cm 4.4cm 10cm},clip]{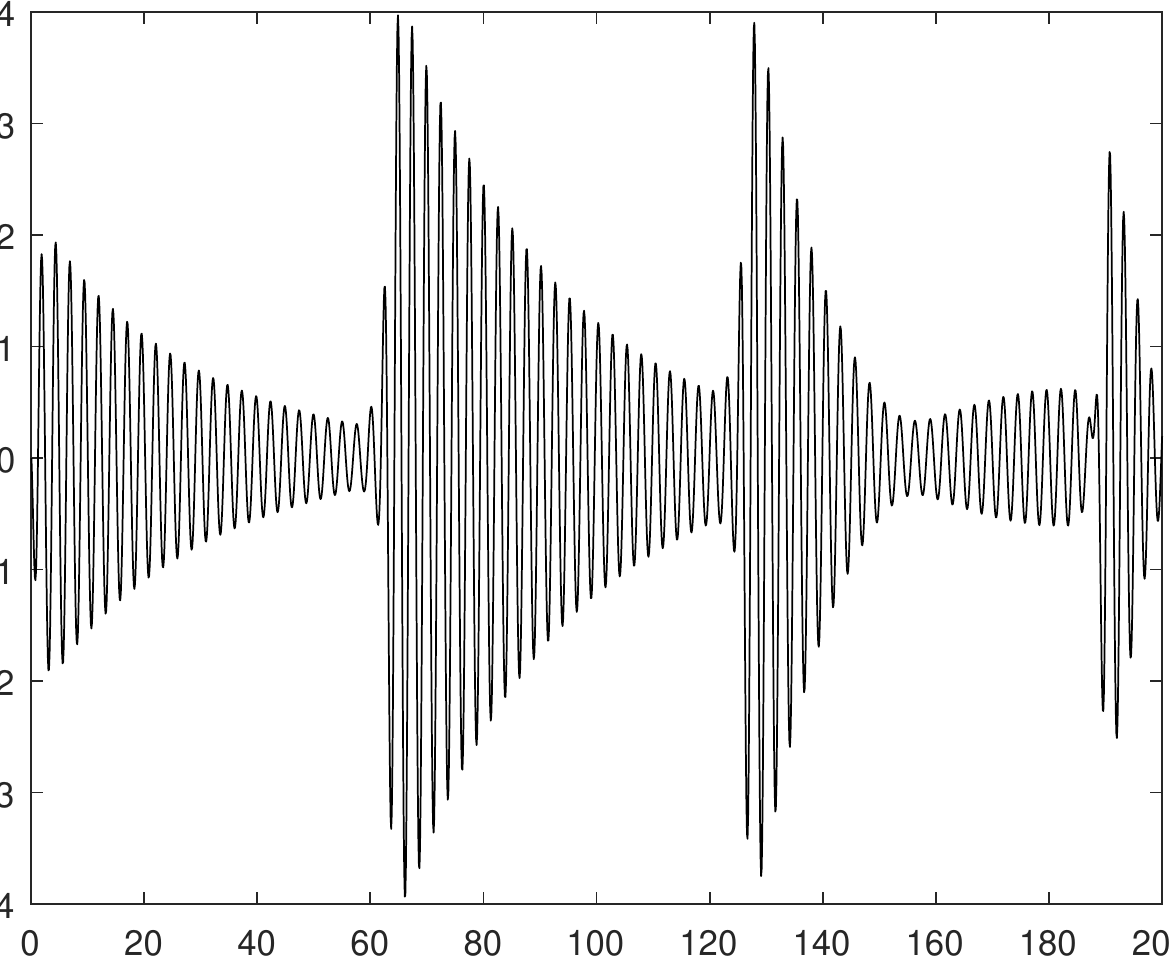}};
  \node[label=left:\rotatebox{90}{Dimensionless position \EX{\Opq}}] at (-4,0) {};
  \node[label=below:{Dimensionless time $\omega_0 t$}] at (0,-3.2) {};
  \node[label={\large \textsf{(a)}}] at (3.2,2.2) {};
\end{tikzpicture}    
\begin{tikzpicture}
        \path (0,0) node(a) {\includegraphics[width=0.9\linewidth,trim={4.5cm 9.4cm 4.3cm 10cm},clip]{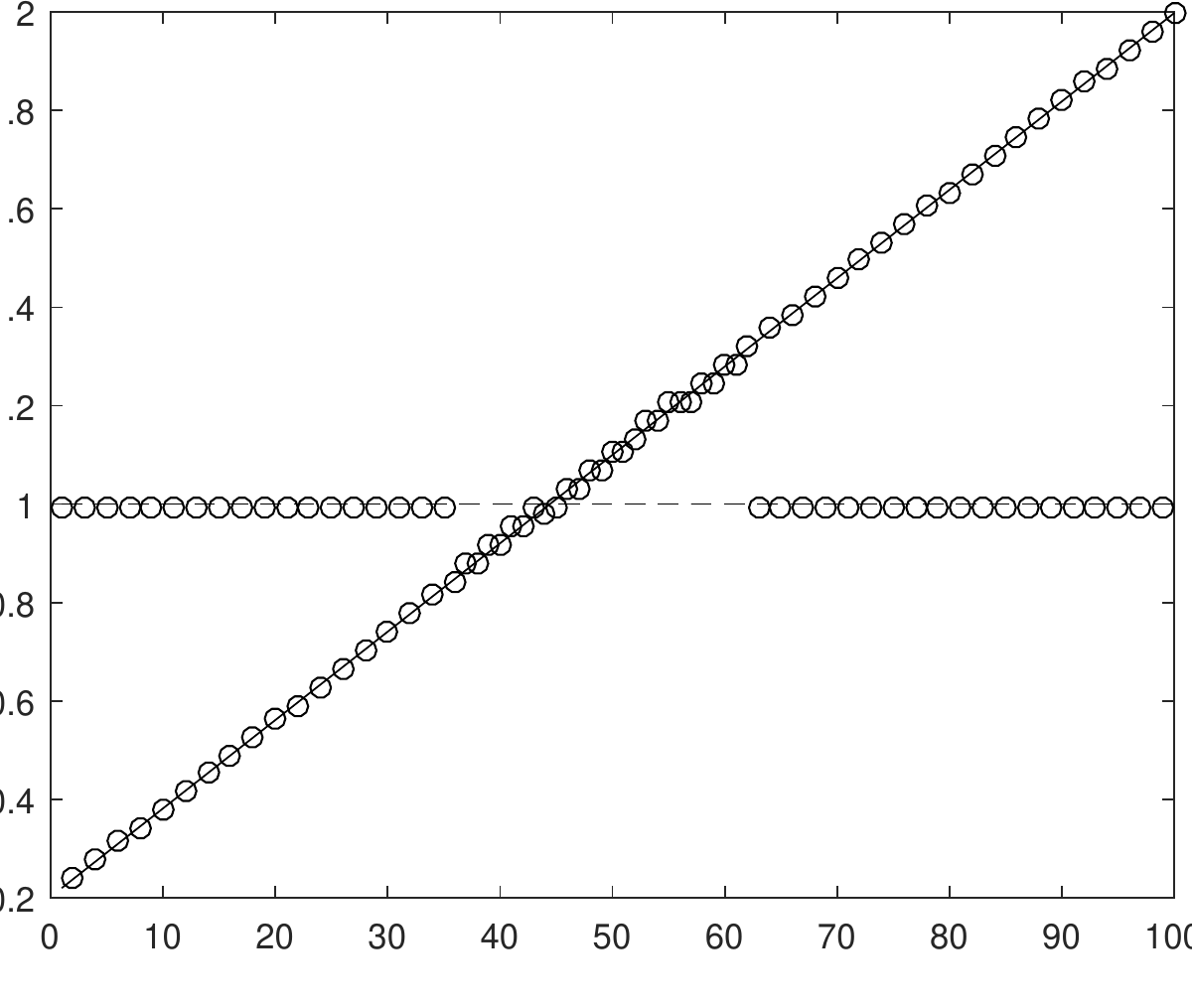}};
  \node[label=left:\rotatebox{90}{Dominant   \EX{\Opa} oscillation frequency$/\omega_0$ }] at (-4,0.2) {};
  \node[label=below:{Bath mode}] at (0,-3.2) {};
  \node[label={\large \textsf{(b)}}] at (3.2,1.8) {};
\end{tikzpicture} 
    \caption{\label{figEntangled} In (a) we show the dynamics of the expectation value of position \EX{\Opq_0} as for~\Fig{figSeperable}(a) but now with the system starting off in the entangled state  \ket{\psi(\xi=1,\zeta=\Half,\delta=\Half)}
as define in~\Eq{eq:psi}. The dynamics of the $j^{th}$ bath mode \EX{\Opq_j} is dominated by oscillation at a frequency shown by the markers in (b). The parameter choices determine the number of modes oscillating at the system frequency $\omega_0$, with the remainder oscillating at their own bath frequency. Linear sampling of the spectral density $J(\omega)$ has been used here after~\cite{Hughes2009}.        }
\end{figure}

Our analysis is not limited to separable initial condition and can readily calculate system dynamics for entangled initial states, a significant problem for Master Equation analysis. Here we use a parametrized initial state that enables us to explore the effects of entanglement. In~\Fig{figEntangled} we show one such example, with the initial state \ket{\psi(\xi=1,\zeta=\Half,\delta=\Half)} defined as 
\bel{eq:psi}
\ket{\psi(\xi,\zeta,\delta)}=\frac{\Big (\frac{\ket{1}+\ket{0}}{\sqrt{2}} \Big )  \otimes \ket {\Psi_1(\xi)} + \delta \Big (\frac{\ket{1}-\ket{0}}{\sqrt{2}}\Big ) \otimes \ket{\Psi_2(\zeta)}}{\sqrt{1+\delta^2}} 
\ee
where
\be
\begin{split}
&\ket{\Psi_1(\xi)}= \Pi_{k=1}^{\otimes N/2} \Big ( \frac{\ket{11}+\ket{00}+\xi ( \ket{01}+\ket{10})}{\sqrt{2(1+\xi^2)}} \Big ) _{2k-1,2k}\\
&\ket{\Psi_2(\zeta)}= \Pi_{k=1}^{\otimes N/2} \Big ( \frac{\ket{11}-\ket{00}+\zeta ( \ket{10}-\ket{01})}{\sqrt{2(1+\zeta^2)}} \Big ) _{2k-1,2k}
\end{split}
\ee
constitute two different ways of entangling the bath modes in pairs $(1,2), (3,4), (5,6), \cdots $. In this initial state $\delta$ allows for control of the degree of initial entanglement between the system and the bath, while $\xi$ and $\zeta$ set an amount of initial entanglement between the various pairs of bath modes. 
The dynamics shown in~\Fig{figEntangled}(a) are very different from those presented for the separable case of~\Fig{figSeperable}(a). Importantly, varying the parameters $\delta$, $\xi$ and $\zeta$ of the initial state has long term impacts on the dynamics of both the system and the bath, synchronizing smaller, or larger, portions of the bath with the system. This may be seen in~\Fig{figEntangled}(b) which shows the location of the peak frequency in the Fourier Transform of the position expectation for each mode. Note that as in~\cite{Hughes2009} the mode frequencies are sampled uniformly from the Spectral Density $J(\omega)$.    

It is interesting to note that, as the temperature of the bath is set by the distribution of photon numbers within each mode, this approach should work equally well in any limit - something that Master Equation models have some difficulty with. Another feature, as discussed above, that this approach has over the Master Equation method is that the system plus environment may be started in an entangled state. This capability opens up new directions for study in quantum thermodynamics, out-of-equilibrium open quantum systems and high temperature quantum phenomena -- making more systems amenable to study, such as indicated in~\cite{PhysRevA.93.062114}.


In this work we have combined the well-known Ehrenfest theorem with our automated symbolic non-commutative algebra package to generate the equations of motion for the expectation values of observables in small to very large ensembles of quantum systems. 
Our methodology demonstrates a new paradigm for the modelling of such quantum systems, enabling exploration, within a fully quantum mechanical framework, of the physics of many systems in a computationally tractable way.
The example considered has an impact already for the study of Quantum Brownian Motion. It is well-known that the universal assumption of a separable initial state is inadequate and leads to ``slips'' in the reduced density matrix dynamics~\cite{Fleming2011}. 
Here we show that, for a system oscillator tuned to a bath with high quality factor, different entangled initial states lead to different dynamics for both the system and the bath. Dependent on the initial state, distinct sections of the bath oscillate freely, whilst others are synchronised to the system frequency.  
Our approach greatly widens not just the range of possible initial states and coupling strengths, but also possible types of coupling and topology as well as for the inclusion of time-dependent Hamiltonians.
As a consequence, it will enable the study of large ensembles of physical systems in a way that was not previously possible.

The Ehrenfest theorem can be used to determine observables such as $\EX{\Opq}(t)$, but also a number of important quantities such as intensity or correlation coefficients including $g^2$ (which may reveal quantum effects such as anti-bunching). It may even be possible to reconstruct the dynamics of the quantum state itself in phase space as the Wigner function (and from this the density matrix) can be written as the expectation value of a displaced parity operator (although momental hierarchy may become an issue here).
Future applications of this method will lead to new insights in the areas of: open quantum systems, emergent behaviour and the quantum to classical crossover, and quantum engineering (including design for test; verification, certification and validation; system design and parameter estimation).



\bibliographystyle{apsrev4-1}
\bibliography{refs}  

\end{document}